\documentclass[12pt]{article}
\usepackage{epsfig}
\usepackage{graphicx}  

\begin{document}
\addtocounter{footnote}{1}
\title{Dark Matter, Dark Energy and the Chaplygin Gas}
\author{Neven Bili\'{c}\thanks{Permanent
 address:
Rudjer Bo\v skovi\'c Institute,
P.O. Box 180, 10002 Zagreb, Croatia;
 \hspace*{5mm} Email: bilic@thphys.irb.hr}\ ,
Gary B.\ Tupper, and Raoul D.\ Viollier\thanks{
 Email: viollier@physci.uct.ac.za}
\\
Institute of Theoretical Physics and Astrophysics, \\
 Department of Physics, University of Cape Town,  \\
 Private Bag, Rondebosch 7701, South Africa \\
 }
\maketitle
\begin{abstract}
We formulate a Zel'dovich-like approximation for the Chaplygin gas
equation of state $P = - A/\rho$, and sketch how this model unifies dark
matter with dark energy in a geometric setting reminiscent of $M$-theory.
\end{abstract}

\section{Introduction}
In the last few years improved observations \cite{mel1} have forced a shift in our
cosmological paradigm:
the $\Omega_{\rm M} = 1 $ dust model has been swept
aside and, in its place, we are faced with the problem of understanding a
universe with an equation of state $\bar{W} = \bar{P}/\bar{\rho} < -
1/3$. That is to say, on average, pressure is comparable with density
and, moreover, negative.

Of course, parametrically,
this is readily accommodated by a cosmological
constant $\Lambda$ \cite{peeb2} with $\Omega_{\Lambda} \simeq$ 0.7 and
$\Omega_{\rm DM} = 1 - \Omega_{\Lambda} \simeq$ 0.3
(throughout the paper we neglect the
small baryonic contribution).
The well-known difficulty with $\Lambda$ is that
{\it a priori} it seems an incredible accident that $\Omega_{\Lambda} \simeq
\Omega_{\rm DM}$ since
$\rho_{\Lambda}/\rho_{\rm DM} \sim a^{3}$, $a$ being the scale
factor.
Hence, much attention has been devoted to quintessence \cite{wett3},
involving a real scalar field which tracks \cite{zlat4} the background
component until recently becoming dominant.
However,
simple tracking quintessence does
not work \cite{blud5}   and spintessence \cite{boyl6}, where the
scalar field is complex,
suffers instabilities against the decay of dark energy
into dark matter \cite{kasu7}.

It is natural to conjecture that some of the aforementioned problems derive
from treating dark matter and dark energy as separate issues. As an example,
Barr and Seckel \cite{barr8} have pointed out that in axion dark matter models
quantum gravity effects break the Pecci-Quinn symmetry leading to a universe
trapped in a false vacuum with an effective $\Lambda$ of the correct
magnitude. In another approach, Wetterich \cite{wett9} has suggested that
traditional WIMP dark matter should
be replaced by quintessence lumps, thus unifying
dark matter and dark energy.
However,
the radiation-matter transition and structure
formation remain  open questions in this scenario.

Herein we present a dark matter-energy unification model
suggested by the observation of Kamenshchik et al. \cite{kamen10} that a
perfect fluid obeying the Chaplygin gas equation of state
\begin{equation}
P = - \frac{A}{\rho}
\label{eq01}
\end{equation}
should lead to a homogenous cosmology with
\begin{equation}
\bar{\rho} (a) = \sqrt{A + \frac{B}{a^{6}}}  ,
\label{eq02}
\end{equation}
with
$B$ being an integration constant, thus interpolating between dark matter,
$\bar{\rho} (a \rightarrow 0) \simeq \sqrt{B}/a^{3}$ and dark energy
$\bar{\rho} (a \rightarrow \infty) \simeq \sqrt{A}$.
Before doing so, we must
first show why Eq. (\ref{eq01}),
aside from its interesting mathematical features
\cite{jach11},
might describe reality.

\section{Brane New World}
One of the most profound recent developments in fundamental physics has been
the recognition that all of
the extra dimensions required by string/$M$-theory
do not have to be of the Planck length size:
one (or more) could be as large as 0.1 mm
provided that all standard-model fields
except gravity are confined to
a 3-dimensional hypersurface or `brane'
in the higher dimensional bulk (for a
review see \cite{ruba12}).
In this context, Kamenshchik et al. \cite{kamen13}
obtained Eq. (\ref{eq01}) from the stabilization of branes in black hole bulks.

A simple way to see the connection
between the Chaplygin gas and the brane
world is to follow Sundrum's \cite{sun14} effective field theory for the 3-
brane.
The gauge fixed embedding of a 3+1 brane in a 4+1 bulk is described by
$Y^{M} = (x^{\mu}, Y^{4})$.
With some nominal assumptions on the bulk metric
$G_{MN}$, the induced metric on the 3-brane is
\begin{equation}
\tilde{g}_{\mu \nu}
=
g_{\mu \nu} - \theta_{, \mu}  \theta_{, \nu} \, ,
\label{eq03}
\end{equation}
with 0 $\leq \theta \leq \ell_{5}$, and the action for the brane reads
\begin{eqnarray}
S_{\rm brane}
 &=&
 \int d^{4} x  \sqrt{- \tilde{g}}
\left[ - f^{4} + \cdots \right]     \nonumber \\
&=&
\int d^{4} x  \sqrt{- g}
\sqrt{1 - g^{\mu \nu} \theta_{, \mu} \theta_{, \nu}}
\left[ - f^{4} - \cdots \right],
\label{eq04}
\end{eqnarray}
where $f^{4}$ is the brane tension and the
ellipsis includes standard-model
fields as well as higher-order terms in power counting.
One estimates
$f \sim \ell_{5}^{-1} \sim$ meV.

Retaining only the leading term in
Eq. (\ref{eq04}) and renaming $f^{4} = \sqrt{A}$,
one sees that its content is equivalent to
\begin{equation}
{\cal{L}} = \frac{\phi^{2}}{2}  g^{\mu \nu}  \theta_{, \mu} \theta_{,
\nu} - V \left( \frac{\phi^{2}}{2} \right)  ,
\label{eq05}
\end{equation}
\begin{equation}
V = \frac{1}{2} \left( \phi^{2} + \frac{A}{\phi^{2}} \right)
\label{eq06}
\end{equation}
since $\phi$ can be eliminated through its field equation
\begin{equation}
g^{\mu \nu} \theta_{, \mu} \theta_{, \nu} = V' \,.
\label{eq07}
\end{equation}
We observe that $\cal{L}$ corresponds to
the Lagrangian for a complex field
$\Phi = \phi e^{-im \theta}/\sqrt{2m}$ in the
`Thomas-Fermi' approximation.
The Thomas-Fermi approximation amounts to
neglecting $\phi_{, \mu}/m \phi$ compared to
$\sqrt{V'/\phi^{2}}$, i.e., the scale of variation of $\phi$ is large
compared to the Compton wavelength.
It is also worth noting that dividing
$\cal{L}$ by $\sqrt{A}$,
the first term is the periodic Gaussian model with
coupling $R = A^{1/4}/\phi$.
The potential $F = (R^{2} + R^{-2})/2$, which is
self-dual, can be interpreted as the mean field free-energy for `brane cells'
filling a system of size $R$, in analogy to Rama's \cite{rama15}
 `string bit'
analysis of black holes.

To complete the connection to the
Chaplygin gas,  we point out a field-fluid
duality:
for $V' > 0$,
Eq. (\ref{eq07}) defines a fluid 4-velocity $U_{\mu} U^{\mu}$ =
1,
\begin{equation}
U^{\mu} = g^{\mu \nu} \theta_{, \nu} / \sqrt{V'}
\label{eq08}
\end{equation}
and then the energy-momentum tensor
derived from $\cal{L}$ takes the perfect
fluid form with
\begin{equation}
\rho = \frac{\phi^{2}}{2}  V' + V  , \hspace{2cm}
P = \frac{\phi^{2}}{2}  V' - V    .
\label{eq09}
\end{equation}
In particular, for $V$ of
Eq. (\ref{eq06}),
the equation of state
 (\ref{eq01}) follows, and the energy density $\rho$
 is given by
\begin{equation}
\rho = \phi^{
2} = \frac{\sqrt{A}}{\sqrt{1-g^{\mu\nu}\theta_{,\mu}
\theta_{,\nu}}}\, ,
\label{eq10}
\end{equation}
which is to say that matter corresponds to a wrinkled brane.

Finally, it must be said that the procedure can be reversed
\cite{bil}.
The equations
\begin{equation}
d \ln  (\phi^{2}) = \frac{d \rho - d P}{\rho + P} \, ,
\hspace{2cm} V = \frac{1}{2} (\rho - P),
\label{eq11}
\end{equation}
obtained from
Eq. (\ref{eq09}), allow one to construct $\phi^{2}$ and $V$ given the
equation of state.
As an example, starting from $0 \geq W = P/ \rho \geq -1$
and  $0 \leq c_{\rm s}^{2} = dP/d \rho \leq 1$,
with the relativistic limits
coinciding, Eqs. (\ref{eq01}), (\ref{eq06}), and (\ref{eq04}) follow.

\section{The Inhomogeneous Chaplygin Gas}
As yet, we have not dealt with the
$\theta$ field equation; in the fluid
language, it reads
\begin{equation}
\left( \sqrt{- g  \phi^{2} (\rho) (\rho + P)} U^{\mu} \right)_{, \mu}
= 0  .
\label{eq12}
\end{equation}
In comoving coordinates, $U^{\mu}=(1/\sqrt{g_{00}},\vec{0})$,
the solution for the Chaplygin gas is
\cite{bil}
\begin{equation}
\rho = \sqrt{A + \frac{B}{\gamma}}  .
\label{eq13}
\end{equation}
Here $\gamma = - g/g_{00}$ is the determinant of the induced metric
$\gamma_{ij} = g_{i0} \, g_{j0}/g_{00} - g_{ij}$ which measures physical
distances, and $B=B(\vec{x})$ can be taken as constant on the
scales of interest.

The generalization
(\ref{eq13}) of Eq. (\ref{eq02}) allows us to implement the
geometric version \cite{matar16}
of the Zel'dovich approximation \cite{zeldo17}:
the transformation from Lagrange to Euler (comoving)
coordinates induces $\gamma_{ij}$ as
\begin{equation}
\gamma_{ij} = \delta_{kl}  {D_{i}}^{k}  {D_{j}}^l\, , \hspace{1cm}
{D_{i}}^{j} = a({\delta_{i}}^{j}-b{\varphi_{,i}}^{j}) \, ,
\label{eq14}
\end{equation}
with ${D_{i}}^{j}$ the deformation tensor, $\varphi$ the velocity potential.
Inserting this ansatz in the 0-0 Einstein equation to first order in $\varphi$
yields the evolution equation for $b(a)$
\begin{equation}
\frac{2}{3}  a^{2} b'' +
a(1 - \bar{w}) b' = (1+\bar{w}) (1-3 \bar{w})b \, ,
\label{eq15}
\end{equation}
\begin{equation}
\bar{w}(a) = - \frac{\Omega_{\Lambda} a^{6}}{1 - \Omega_{\Lambda} +
\Omega_{\Lambda} a^{6}}\, ,
\label{eq16}
\end{equation}
where we match the parameters $A$, $B$ to the
$\Lambda$ model.

\begin{figure}[h]
\begin{center}
\includegraphics[width=.6\textwidth]{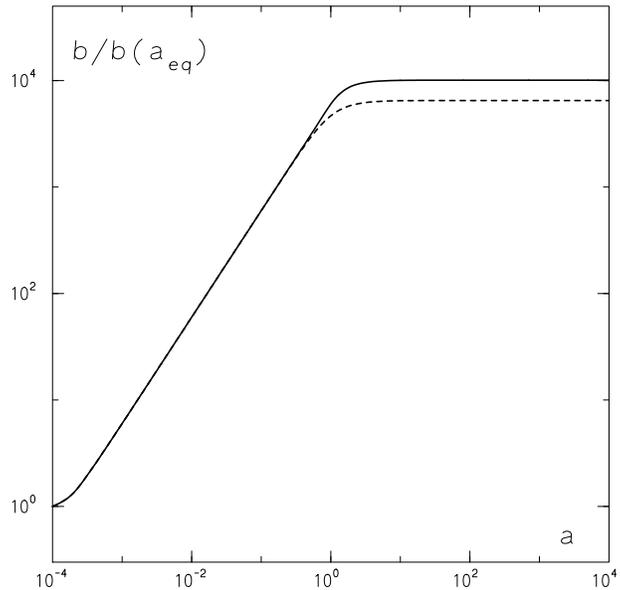}
\end{center}
\caption{
 Evolution of $b (a)/ b(a_{\rm eq})$
from $a_{\rm eq}= 1.0 \times 10^{-4}$ for $\Omega_{\Lambda} = 0.7$
    and $b' (a_{\rm eq}) = 0$,
   for the Chaplygin gas (solid line) and $\Lambda$CDM (dashed line).
}
\label{fig}
\end{figure}
In Fig. \ref{fig} we
show the evolution of $b(a)$ for the Chaplygin gas and for $\Lambda$ cold dark
matter, the latter following by omitting the $(1 - 3 \bar{w})$ factor in
Eq. (\ref{eq15}) and changing
$a^{6}$ to $a^{3}$ in Eq. (\ref{eq16}). In either case, the growth
$b \propto a$ ceases near $a = 1$ and although $b$
remains constant,
the perturbative density contrast $\delta_{\rm pert} = b(1+ \bar{w})
{\varphi,_{i}}^{i}$
thereafter vanishes as $\delta_{\rm pert} (a \gg 1) \sim a^{-
6}$.

Of course the value of the
Zel'dovich approximation is that it offers a means
of extrapolation into the
nonperturbative regime via Eqs. (\ref{eq13}) and
\begin{equation}
\sqrt{\gamma} = a^{3}
(1-\lambda_{1}b)(1-\lambda_{2}b)(1-\lambda_{3}b),
\label{eq17}
\end{equation}
where the $\lambda_{i}$ are the eigenvalues of
${\varphi_{,i}}^{j}$.
When one
(or more) of the $\lambda$'s is positive,
a caustic forms on which $\gamma
\rightarrow 0$ and $w \rightarrow 0$, i.e.,
at the locations where structure
forms the Chaplygin gas behaves as dark matter.
Conversely, when all of the
$\lambda$'s are negative, a void forms,
$\rho$ is driven to its limiting value
$\sqrt{A}$, and the Chaplygin gas
behaves as dark energy driving accelerated
expansion.

\section{Discussion and Conclusions}
A shortcoming of the Zel'dovich approximation is that at the caustic matter
flows through unimpeded so that structures quickly dissolve \cite{pauls18}.
This may be circumvented via the truncated Zel'dovich approximation
\cite{pauls18}.
A preferable alternative would be an extension of the adhesion
approximation \cite{gurb19} which
also
allows the extraction of mass functions.

Approximation technicalities aside, the case is made that the Chaplygin gas
offers a realistic unified model of dark matter and dark energy.
That this is
achieved in a geometric
(brane world) setting rooted in string/$M$ theory
makes this model all the more remarkable.


\end{document}